\documentclass[aps,prb,twocolumn]{revtex4}
\usepackage{amsfonts}
\usepackage{amsmath}
\usepackage{amssymb}
\usepackage{graphicx}

\begin{document}


\title{Spin-dependent  transmission through a chain of rings:
influence of a periodically modulated spin-orbit interaction
strength or ring radius}
\author{B. Moln\'{a}r}
 \affiliation{Departement Natuurkunde, Universiteit Antwerpen
(Campus Drie Eiken), B-2610 Antwerpen, Belgium}

\author{P. Vasilopoulos}
\altaffiliation{Department of Physics, Concordia University, 1455
de Maisonneuve Quest, Montr\'eal, Quebec, Canada H3G 1M8}

\author{F. M. Peeters}
\affiliation{Departement Natuurkunde, Universiteit Antwerpen
(Campus Drie Eiken), B-2610 Antwerpen, Belgium}

\begin{abstract}
We  study ballistic electron transport through a finite chain of
quantum circular rings in the presence of  spin-orbit interaction
of strength $\alpha$. For a single ring the transmission and
reflection coefficients  are obtained analytically and from them
the conductance for a  chain  of  rings as a function of $\alpha$
and of the wave vector $k$ of the incident electron. We show that
due to destructive spin interferences the chain can be totaly
opaque for certain ranges of $k$ the width of which depends on the
value of $\alpha$. A periodic modulation of the strength $\alpha$
or of the ring radius widens up the gaps considerably and produces
a nearly binary conductance output.




\end{abstract}

\maketitle

In recent years the study of spintronics devices, which utilize
the spin rather than the charge of an electron, has been
intensified mainly because they are expected to operate at much
higher speeds than the conventional ones and have potential
applications in quantum computing. One such device is a single
ring in the presence of the Rashba coupling \cite{1} or spin-orbit
interaction (SOI) which results from asymmetric confinement in
semiconductor nanostructures. It is  important in materials with a
small band gap such as InGaAs. An important feature of  electron
transport through a ring is that, even in the absence of an
external magnetic field, the difference in the Aharonov-Casher
phase of carriers travelling clockwise and counterclockwise
produces spin-sensitive interference effects.\cite{2} Here we
build on this fact by studying electron transport through a chain
of identical rings in the presence of SOI of strength $\alpha$. In
addition, we study the influence of periodic modulations of
$\alpha$ or of the ring radius.
The main motivation behind this study is to produce a
controllable, binary output, with wide gaps, upon periodically
varying $\alpha$ and/or ring radius. Such binary output is
pertinent to the development of the spin transistor.\cite{3}
Here are the context and the results.

{\bf Single-ring transport.} In the presence of SOI the
Hamiltonian operator for a  one-dimensional ring reads
\cite{4}
\begin{eqnarray}
H=\hbar \Omega 
\big[ -i\frac{\partial}{\partial \varphi}
+\frac{\omega _{so}}{2\Omega}(\cos \varphi \sigma _{x}+\sin
\varphi \sigma _{y})\big]
^2 \label{Ha1}
\end{eqnarray}
where $\sigma _{x}$, $\sigma _{y}$, and $\sigma _{z}$ are the
Pauli matrices. The parameter $\Omega $ denotes $\hbar /2m^{\ast
}a^{2}$ and $\omega _{so}=\alpha /\hbar a$ is the frequency
associated with the SOI. The parameter $\alpha $ represents the
average electric field along the $z$ direction. For an
InGaAs-based system $\alpha$ can be controlled by a gate voltage
with values in the range $(0.5-2.0)\times
10^{-11}\mathrm{eVm}$.\cite{5}

{\it Eigenvalues, eigenfunctions.} The energy spectrum $E_{n}^{(\mu )}$ and
unnormalized eigenstates $\Psi _{n}^{(\mu )}$ pertaining to Eq. (1) are
labeled by the index $\mu
=1,2$ and given by \cite{6}
\begin{equation}
E_{n}^{(\mu )} =\hbar\Omega(n-\Phi _{AC}^{(\mu )}/2\pi )^{2}, \\\
\Psi _{n}^{(\mu )}(\varphi ) =e^{in\varphi }\chi^{(\mu )}(\varphi
); \label{stat}
\end{equation}
the orthogonal spinors $\chi ^{(\mu )}(\varphi )$ can be expressed
in terms of the eigenvectors $[1,0]^T$, $[0,1]^T$ of the  matrix
$\sigma _{z}$ as
\begin{equation}
\hspace{-0.00cm} \chi^{(1)}= [\cos \frac{\theta}{2}, e^{i\varphi
}\sin\frac{\theta }{2}]^T, \chi^{(2)}=[\sin \frac{\theta}{2},
-e^{i\varphi }\cos \frac{\theta }{2}]^T \label{spin2}
\end{equation}%
with $\chi^{(\mu)}\equiv \chi^{(\mu)}(\varphi )$, $T$ denoting the
transpose of the row vectors, and $\theta =2\arctan[ \Omega/\omega
_{so} -(\Omega ^{2}/\omega _{so}^{2}+1)^{1/2}]. $
 The spin-dependent term $\Phi _{AC}^{(\mu )}$ is the
Aharonov-Casher phase $\Phi _{AC}^{(\mu )}=-\pi [ 1+(-1)^{\mu }(
\omega _{so}^{2}+\Omega ^{2}) ^{1/2}/\Omega ] $.

The ring connected to two leads is shown in the inset to Fig. 1
with the local coordinate systems attached to the different
regions. The appropriate boundary conditions to apply at the
intersections are a spin-dependent version of Griffith's boundary
conditions.\cite{6} Specifically, at each junction the wave
function must be continuous and the spin probability current
density must be conserved.

In the present problem the total wave function in the lead
can be expanded in terms of spinors
$\chi ^{(\mu )}$.
We have
\begin{eqnarray}
\Psi _{I}(x)& =&\underset{\mu =1,2}{\sum }[e^{ikx}f^{(\mu)
}+e^{-ikx}r^{(\mu) }]\chi ^{(\mu )}(\pi ), \\* \Psi
_{II}(x^{\prime })& =&\underset{\mu =1,2}{\sum
}(-1)^{\mu+1}[e^{ikx^{\prime }}t^{(\mu) }+e^{-ikx^{\prime
}}g^{(\mu) }]\chi ^{(\mu )}(0),
\end{eqnarray}
in region I and II, respectively, and $k$ denotes the incident
wave vector. The coefficients $f^{(\mu)}$ ($g^{(\mu)}$) are the
amplitudes of the spin state $\mu=1,2$ for electrons incident from
the left (right) lead and $r^{(\mu)}$ ($t^{(\mu)}$) those of the
spin state which are reflected to the left (exiting to the right)
of the ring.
One can show that the spinor $\chi ^{(\mu )}(0)=(-1)^{\mu+1}U \chi
^{(\mu )}(\pi)$, with the unitary operator $U$ having the form $
U=\left[
\begin{array}{cc}
 \cos \theta & - \sin \theta \\
 \sin \theta & \cos \theta
\end{array}
\right]. $
A similar expansion can be made for the wave functions in the upper
and lower arms of the ring.
The result is
\begin{eqnarray}
\Psi _{up}(\varphi )& =&\overset{2}{\underset{\mu,j=1}{\sum
}}a_{j}^{(\mu) }e^{in_{j}^{\mu }\varphi }\chi ^{(\mu )}(\varphi),
\\*
\Psi _{low}(\varphi ^{\prime })& =&\overset{2}{\underset{\mu,j=1}{\sum }}%
b_{j}^{(\mu) }e^{-in_{j}^{\mu }\varphi ^{\prime }}\chi ^{(\mu )}(-
\varphi^{\prime}),
\end{eqnarray}
with $n_{j}^{\mu }=(-1)^{j}ka+\Phi _{AC}^{(\mu )}/2\pi $ the
solutions of the equation $k^{2}a^{2}=E_{n^{\mu }}^{\mu
}/\hbar\Omega$ that ensure energy conservation.

{\it Reflection and transmission coefficients.} Applying the
boundary conditions one can verify that the amplitudes $g^{(\mu)
}$ and $t^{(\mu) }$ are connected to $r^{(\mu) }$ and $f^{(\mu) }$
by a transfer matrix $L$ independent of the spin index $\mu$ as
\begin{equation}
L[r^{(\mu)} , f^{(\mu)}]^T =[g^{(\mu)},t^{(\mu)}]^T.
\end{equation}
The matrix $L$ above can be written in an analytic form
\begin{equation}
L=(1/T)\left[
\begin{array}{cc}
1 & \ \ -R\\
R & \ \ (T)^2 -(R)^2
\end{array}
\right],
\end{equation}
where $T$ and  $R$ are the functions of $ka$ and
$\Delta_{AC}=(\Phi^{(1)}_{AC}-\Phi^{(2)}_{AC})/2$ (hence of SOI
strength $\alpha$)
\begin{eqnarray}
T=\frac{i\sin(\Delta_{AC}/2)\sin (ka\pi )}{\sin^2( \Delta
_{AC}/2)-[\cos (ka\pi )-i\sin (ka\pi )/2]^2},
\label{ftrans} \\
R=\frac{[1+3\cos(2ka\pi)+4\cos\Delta _{AC}]/8}{\sin^2( \Delta
_{AC}/2)-[\cos (ka\pi )-i\sin (ka\pi )/2]^2}.
\end{eqnarray}
Let us assume that an electron enters the ring from the left with
an arbitrary spin orientation ($f^{(\mu)}$ are arbitrary complex
numbers) but $g^{(1)}=g^{(2)}=0$, i. e., that there is no incident
electron current from the right. Then the electron is reflected
without changing its original spin-orientation with the
probability amplitude $R=-L_{12}/L_{11}$; on the other hand, it is
transmitted with the transmission coefficent
$T=-L_{12}L_{21}/L_{11}+L_{22}$ but its spin is unitarily rotated
by $U$. One can show that the standard relation $\vert T\vert
^{2}+\vert R\vert ^{2}=1$ is held.

{\bf Multi-ring conductance.} If we have $N$ rings, the
single-ring result can be easily generalized for a chain if the
rings  only touch each other, cf. inset of Fig. 3(d1). First, one
has to calculate the joint transfer matrix $\tilde{L}$
\begin{equation}
\left[
\begin{array}{c}g^{(\mu)}_{N}\\t^{(\mu)}_{N}\end{array}
\right]=\tilde{L}\left[
\begin{array}{c}r^{(\mu)}_{1}\\f^{(\mu)}_{1}\end{array}
\right]=L_{N}...L_{1}\left[
\begin{array}{c}r^{(\mu)}_{1}\\f^{(\mu)}_{1}\end{array}
\right],
\end{equation}
then apply the boundary condition $g^{\mu}_{N}=0$, that is, after
the last ring there is only outgoing wave function. Then the
reflection ($\widetilde{R}$) and transmission ($\widetilde{T}$)
coefficients are written in terms of $\tilde{L}$ and the
conductance $G$ reads
\begin{equation}
G=2(e^2/h)\left \vert \widetilde{T}\right \vert
^{2}=2(e^2/h)\left\vert \tilde{L}_{12}\tilde{L}_{21}/
\tilde{L}_{11}-\tilde{L}_{22}\right \vert ^{2}.
\end{equation}
We note that the spin of the electrons exiting the chain is
rotated by $\tilde{U}=U(N)..U(1)$ in respect to that of the
incident electrons.

{\bf Numerical results.} In Fig. 1 the conductance $G(ka)$,
through a chain of $N=101$ rings, is shown as a function of the
incident wave vector $ka$ for various values of $\alpha$. Because
$G(ka)$ is an even and periodic  function of $ka$ with period 1 we
show it only within one period of $ka$, for $5\leq ka\leq 6$. In
the absence of SOI the conductance $G$ oscillates with high values
but it never drops to zero. In other words, a chain without SOI is
\textit{never} totally reflexive. But for finite non zero values
of $\alpha$ it  takes a "square-wave"  form and has \textit{zero}
value in a \textit{finite} range of $k$, the width of which
strongly depends on  $\alpha$. Outside these ranges $G$  always
oscillates with high values.

\begin{figure}
  \includegraphics[width=5cm,angle=-90]{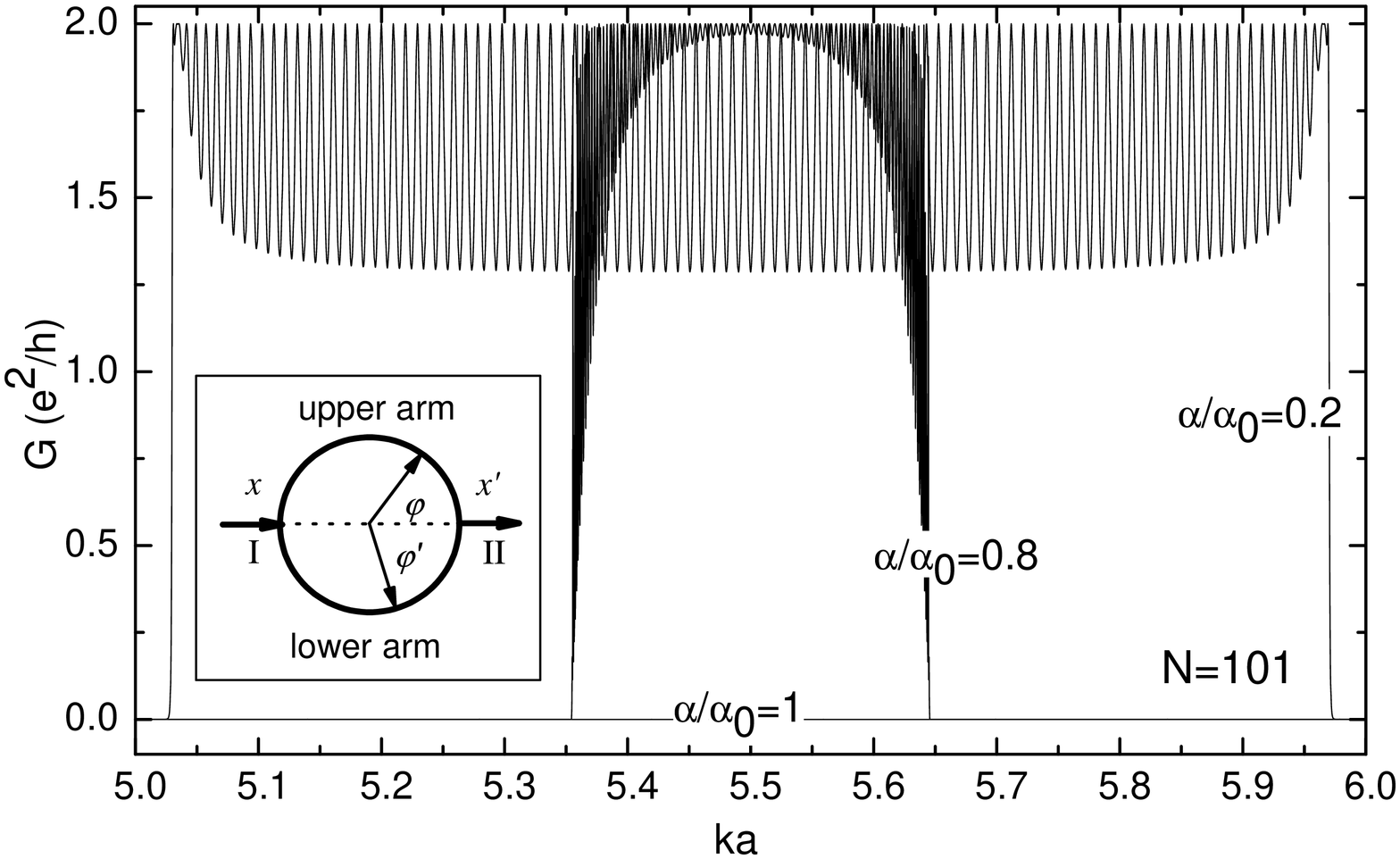}\\
    \caption{Conductance through a chain of $N=101$ rings versus $ka$
for some values of $\alpha$
($\alpha_{0}=1.147\times10^{-11}\mathrm{eVm}$). The inset shows
the local coordinates ($x$, $x^{\prime}$, $\varphi$, and
$\varphi^{\prime}$) pertaining to different regions of the ring.}
\end{figure}

The occurence of the gaps in $G(ka)$ is
attributed to the Aharonov-Casher effect due to the spin
precession induced by the SOI and the destructive interference
between the electron spins travelling in the clockwise and
counter-clockwise directions. 
In the most extreme cases, if the SOI strength has certain
well-defined values
$\alpha=(\hbar^{2}/2m^{\ast}a)\sqrt{4(n+1)^{2}-1}\equiv\alpha_{n}$,
for $n$ an arbitrary integer, the difference between
$\Phi^{(1)}_{AC}=(2n+1)\pi$ and
$\Phi^{(2)}_{AC}=-2\pi-\Phi^{(1)}_{AC}=-(2n+3)\pi$ renders the
spin interference destructive and leads to the widest gap because
each ring is non-transparent for any value of $ka$. With the
effective mass of InAs $m^{\ast}=0.023m_{0}$ and a ring radius
$a=0.25\mathrm{\mu m}$ the smallest value ($n=0$) of $\alpha$
which can  produce total reflection  in the chain (and used for
the results shown) is
$\alpha_{0}=1.147\times10^{-11}\mathrm{eVm}$.

\begin{figure}
  \includegraphics[width=5cm,angle=-90]{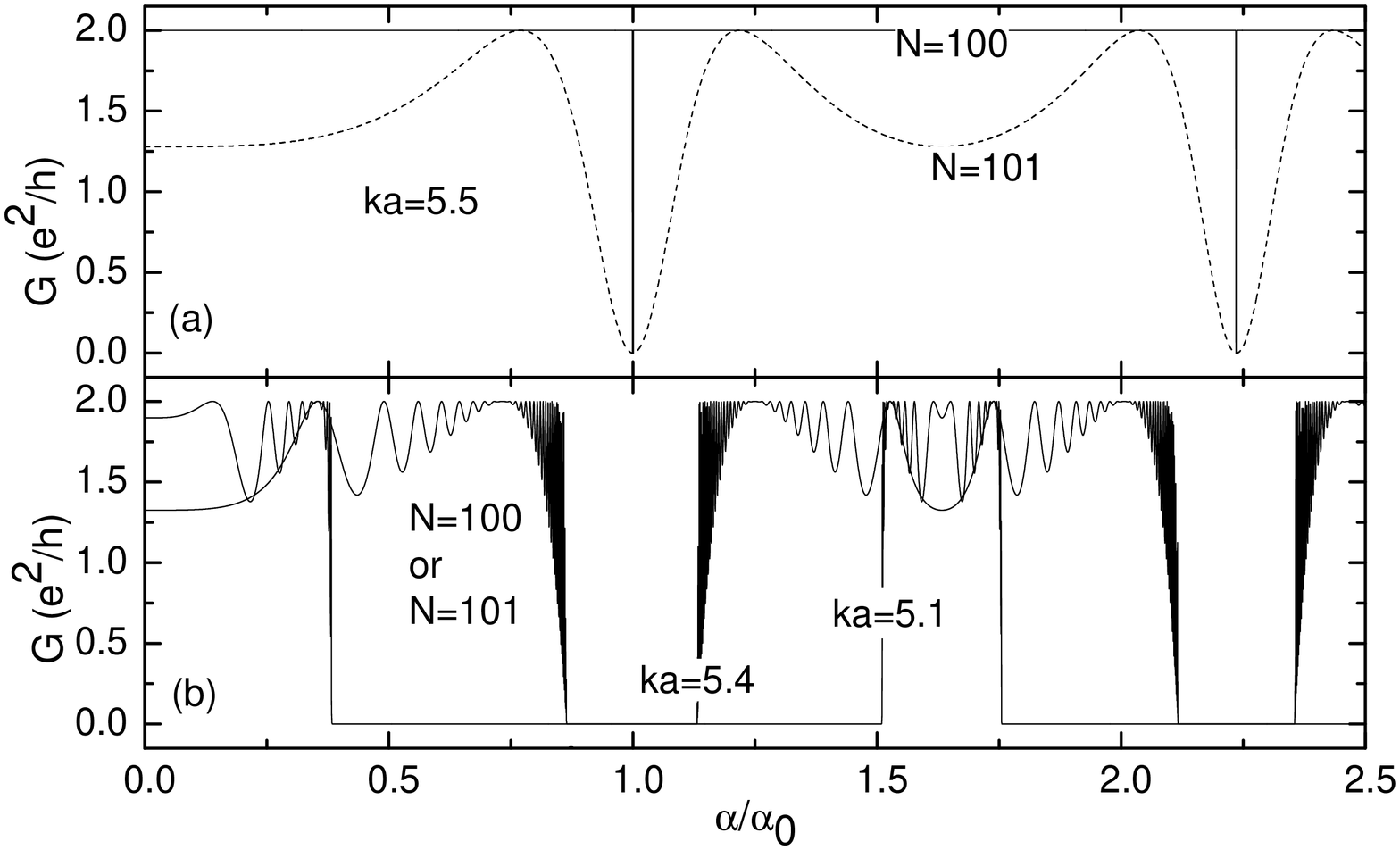}\\
    \caption{Figure 2: Conductance versus $\alpha$ for various $ka$ through a
chain of $N=100$ and $101$ rings.}
\end{figure}

The rapid oscillations and the square profile of $G(k)$ stem from
the fact the chain contains many identical rings. In general,  the
low transmission values for a single ring become almost zero for a
chain with many rings while those values near the maximum 1 remain
nearly unchanged. However, this is not exactly true if $ka$ is a
half integer. For such a $ka$ we have
$\tilde{L}\tilde{L}=-\hat{I}$, where $\hat{I}$ is the identity
matrix; consequently, the conductance of a finite chain depends on
the parity of $N$. More precisely, if $N$ is odd $G$ equals the
single ring conductance, while if $N$ is even $G$ is a
discontinuous function which vanishes only at $\alpha
=\alpha_{n}$, $n$ integer, and otherwise has the value 2. If $ka$
is relatively far from a half integer no difference can be seen
between $N$ and $N+1$ for $N=100$, cf. Figs. 2(a) and 2(b).

\begin{figure}
  \includegraphics[width=5cm,angle=-90]{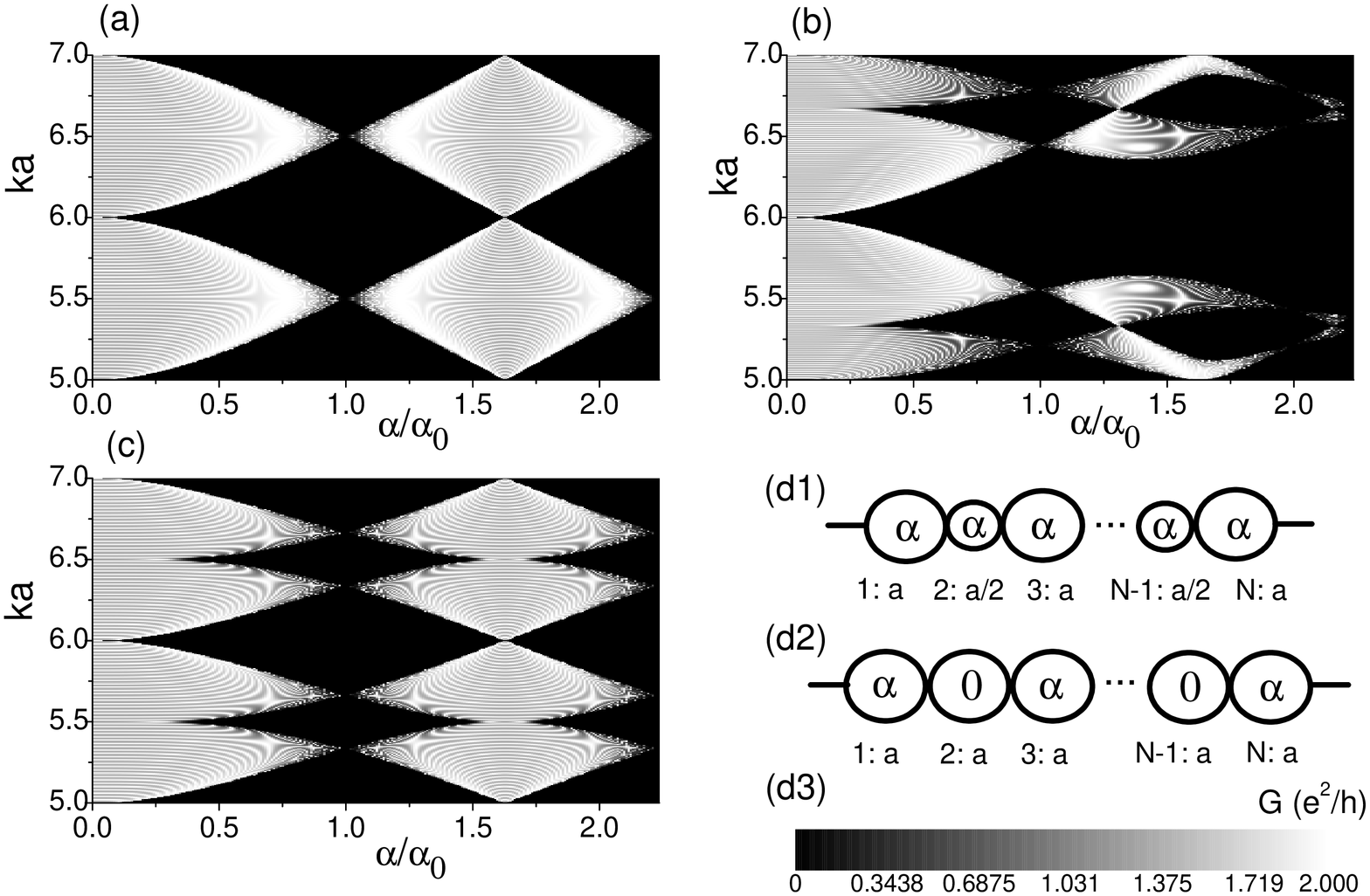}\\
    \caption{ Contour plot of the conductance through a chain of
$N=101$ rings versus $ka$ and $\alpha$. In (a) $a$ and $\alpha$
are fixed, in (b) $\alpha$ is kept fixed but $a$ changes as shown
in (d1), and in (c) $a$ is fixed while $\alpha$ changes as shown
in (d2). Panel (d3) shows the grey color intensity.}
\end{figure}

Fig. 3(a) is a grey-scale contour plot summarizing the behavior of
the conductance $G$ versus  $ka$ and $\alpha$. The darkest regions
correspond to ranges of $ka$ and $\alpha$ for which  the chain is
opaque ($G=0$) while the white ones correspond to those for which
the maximum conductance value is $2e^{2}/h$ and the chain is
transparent. Figure 3(b)  shows how a periodic modulation in the
ring radius modifies the conductance profile of Fig. 3(a). The
radius of the $i^{\mathrm{th}}$ ring is given by $a_{i}=a$, if $i$
is odd, and $a_{i}=a/2$, if $i$ is even. As seen, the large bright
regions  along  the lines $ka\approx5.3$ and $6.7$,  with high
conductance, are split apart and produce a full, wide gap along
the line $ka=6$ for a wider range of $\alpha$. Also,  several new
gaps appear.

\begin{figure}
  \includegraphics[width=5cm,angle=-90]{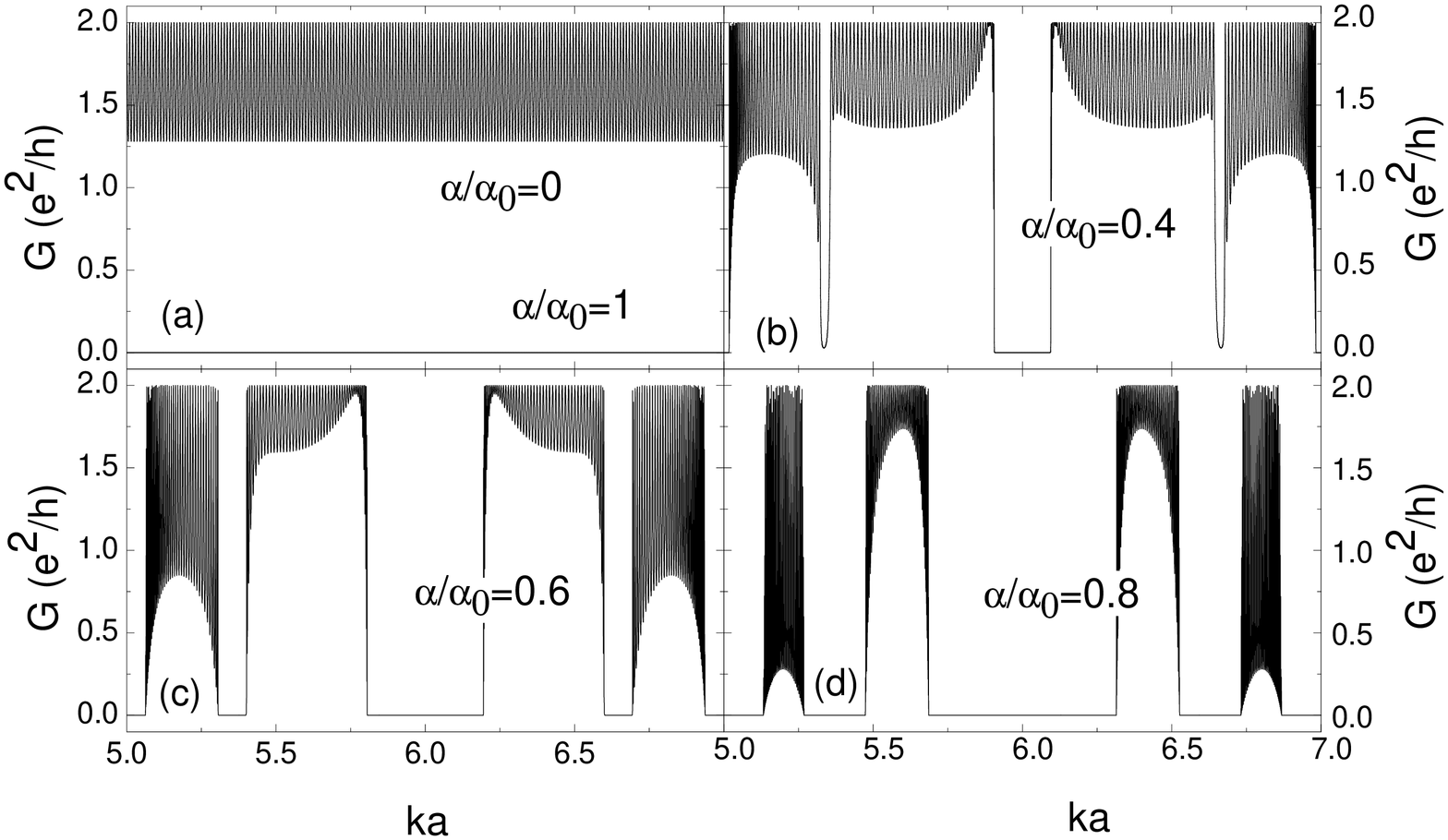}\\
    \caption{Conductance versus $ka$  through a chain  of $N=101$
rings pertaining to Fig. 3(c) and the indicated  values of
$\alpha$.}
\end{figure}

A more regular behavior can be obtained if only $\alpha$ is
modulated  and the radius is kept fixed. For a chain of $N=101$
rings having the same radius $a$ but with SOI $\alpha$ changing
from ring to ring as $\alpha_{i}=\alpha$, if $i$ is odd, and
$\alpha_{i}=0$, if $i$ is even, the resulting $G(k,\alpha)$ is
shown in Fig. 3(c). Such a profile for $\alpha$ could be created
by  applying a gate to only to the $i$-th ring, $i$ for odd. The
contour plot is more symmetric and has definite dark gaps as in
the case of a single waveguide.\cite{7} Relative to Fig. 3(a) we
see more gaps for constant $\alpha$ and variable $ka$. Another way
of appreciating the results of Fig. 3(c) is shown in Fig. 4 in
which $G(k,\alpha)$ is plotted along the lines of Fig. 3(c) at
$\alpha/\alpha_{0}=0$, $0.4$, $0.6$, $0.8$ and $1$. Notice that in
each panel $\alpha$ changes from ring to ring as  shown in Fig.
3(d2). As for the widening of the gaps conductance gaps in Figs.
3(c) and Fig. 4 relative to those in Fig. 3(a), it can be
understood qualitatively with the help of Eq. (2): upon
periodically varying the ring radius or strength $\alpha$,
$\omega_{so}$ and the energy levels change from ring to ring thus
creating the usual superlattice barriers or wells. Depending then
on the incident electron's energy one has the usual gaps or bands
in the transmission.

The results presented here are valid for chains of strictly
one-dimensional rings. They can be extended to rings of finite
width $w$ provided the inequality $w \ll a$ holds and, e.g., an infinite
well confinement is assumed along the radial direction. In this
case the radial and angular motion are decoupled and the energy
levels, given by Eq. (2), are shifted by
$\hbar^{2}l^{2}/2m^{\ast}w^{2}$, where $l$ is an integer. The
results presented above correspond then to the lowest $l=1$ mode.

In summary, we studied  ballistic electron transport   through
chains of rings in the presence of  SOI, of strength $\alpha$, and
showed that gaps in the conductance, as a function of $\alpha$
and/or  the electron's wave vector $k$, occur due to destructive
interferences between  electron spins travelling in the clockwise
and counter-clockwise directions. In particular, we showed that
periodic modulations of $\alpha$ or of the ring radius  widen
these gaps and produce a nearly square-wave conductance.
The full gaps in the conductance plotted in Figs. 2-4 occur
whether the incident electrons are spin polarized or not.
Accordingly, the results are pertinent to the development of the
spin transistor where a spin-dependent and binary conductance
output is necessary with as a good control as possible.

This work was supported by the Belgian Interuniversity Attraction
Poles (IUAP), the Flemish Concerted Action (GOA) Programme, the
Flemish Science Foundation (FWO-Vl), the EU-CERION programme, the
Flemish-Hungarian Bilateral Programme and by the Canadian NSERC
Grant No. OGP0121756. One of us (B. M.) is supported by DWTC to
promote S \& T collaboration between Central and Eastern Europe.\












\begin{thebibliography}{99}
\bibitem{1} E. I. Rashba, Sov. Phys. Solid State \textbf{2}, 1109
(1960).
\bibitem{2} Y. Aharonov and A. Casher, Phys. Rev. Lett. \textbf{53}, 319
(1984);  A. G. Aronov and Y. B. Lyanda-Geller, {\it ibid}
\textbf{70}, 343 (1993); S. L. Zhu and Z. D. Wang, {\it ibid}
\textbf{85}, 1076 (2000);  D. Frustaglia, M. Hentschel, and K.
Richter, {\it ibid} \textbf{87}, 256602 (2001); J. B. Yau, E. P.
De Poortere, and M. Shayegan, {\it ibid} \textbf{88}, 146801
(2003).

\bibitem{2a} S. Datta and B. Das, Appl. Phys. Lett. \textbf{56},
665 (1990).
\bibitem{4} F. E. Meijer, A. F. Morpurgo, and T. M. Klapwijk, Phys. Rev. B \textbf{66}, 033107 (2002).

\bibitem{5} D. Grundler, Phys. Rev. Lett. \textbf{84}, 6074 (2000).

\bibitem{6} B. Molnar, F. M. Peeters, and P. Vasilopoulos, Phys. Rev. B \textbf{69}, 155335 (2004).  A constant term $\hbar\omega_{so}^2/4\Omega$
is neglected in the expression for the eigenvalues and in Eq. (1). The neglect is acceptable for $ka\gg \omega_{so}/2\Omega$. 

\bibitem{7} J. B. Xia, Phys. Rev. B \textbf{45}, 3593 (1992); T. Choi, S. Y. Cho, C. M. Ryu, and C. K. Kim, {\it ibid}
\textbf{56}, 4825 (1997).

\bibitem{8} X. F. Wang and  P. Vasilopoulos, Appl. Phys. Lett. \textbf{82}, 940 (2003).
 \end{thebibliography}
\end{document}